\documentclass[%
 aip,
 amsmath,amssymb,
 reprint,%
]{revtex4-1}

\usepackage{graphicx}
\usepackage{dcolumn}
\usepackage{bm}

\usepackage[utf8]{inputenc}
\usepackage[T1]{fontenc}
\usepackage{mathptmx}
\usepackage{mathtools}

\usepackage{natbib}
\usepackage{amsmath}
\usepackage{amsfonts}
\usepackage{newtxmath}

\begin{document}

\preprint{AIP/123-QED}

\title[]{Inhomogeneous surface tension of chemically active fluid interfaces}

\author{Alessio Squarcini}
\email{squarcio@is.mpg.de}
\author{Paolo Malgaretti}
\email{malgaretti@is.mpg.de}
\affiliation{Max-Planck-Institut f{\"u}r Intelligente Systeme, Heisenbergstr.~3, D-70569 Stuttgart, Germany}
\affiliation{IV Institut f\"ur Theoretische  Physik, Universit\"at Stuttgart, Pfaffenwaldring 57, D-70569 Stuttgart, Germany}

\date{\today}

\begin{abstract}
We study the dependence of the surface tension of a fluid interface on the density profile of a third suspended phase. By means of an approximated model for the binary mixture and of a perturbative approach we derive close formulas for the free energy of the system and for the surface tension of the interface. Our results show  a remarkable non-monotonous dependence of the surface tension on the peak of the density of the suspended phase. Our results also predict the  local value of the surface tension in the case in which the density of the suspended phase is not homogeneous along the interface. 
\end{abstract}

\maketitle

\section{Introduction}
The physics of liquid interfaces is crucial in several scenarios~\cite{Fuller2012} including the stability of biofilms in healthcare devices~\cite{Lindsay2006}, shipping~\cite{Flemming2006}, self-cleaning surfaces for automotive industry~\cite{B602486F}.
Clearly, the mechanical and thermodynamic properties of the interface are ultimately determined by the concentrations and interactions among its molecular constituents. 
It is well known that adding additional suspended particles to a phase-separated binary mixture can alter the interface and in particular it surface tension.
Interestingly, in many scenarios of actual interest, fluid interfaces are either formed by complex fluids whose structure may change in time, as it happens for antagonistic salts~\cite{Harting2019,Glende2020} or they are in the vicinity of  ``devices'' that continuously transform some molecules into others. This is the case of biofilm formation~\cite{Lindsay2006,C9SM02038A}, where the cellular activity keeps on pumping molecules in and out the cell membrane. Similarly, phoretic colloids attains motion by catalyzing chemical reactions on their surfaces. 
When such system are in the vicinity of a fluid interface~\cite{Malgaretti2016,Peter2020}, their chemical activity affects the local surface tension possibly leading to a variety of phenomena including the onset of Marangoni flows~\cite{Dominguez2016}.
 
In this contribution, we determine the dependence of the surface tension of a phase-separated binary mixture on the density of a third suspended phase.
We show that the presence of this additional suspended phase affects the surface tension. In particular, when the concentration of the suspended phase is not homogeneous, the surface tension varies along the interface. We show that when the density of the suspended phase is not at equilibrium (i.e. its chemical potential is not homogeneous) the imbalance in the surface tension is not compensated for by any conservative force and hence it leads to the onset of Marangoni flows. At variance, when the suspended phase experience an external conservative potential, the system still reaches an equilibrium steady state characterized by a homogeneous chemical potential and an inhomogeneous density profile. 
In such a scenario, surface tension is still inhomogenous but the force induced by this inhomogeneity will be compensated for by the conservative potential hence leading to no Marangoni flows. 
The structure of the manuscript is as follows. In Sec.~\ref{sec:model} we describe the coarse-grained model of the binary mixture and its interaction with the suspended phase. In Sec.~\ref{sec3} we compute the density profile of the suspended phase, in Sec.~\ref{sec4} we derive the protocol to calculate the surface tension and we discuss some interesting scenario. Finally in Sec.~\ref{sec5} we derive the expression for the local surface tension and in Sec.~\ref{sec:concl} we draw our conclusions.
\begin{figure}[htbp]
\begin{center}
\centerline{\includegraphics[width = .5 \textwidth]{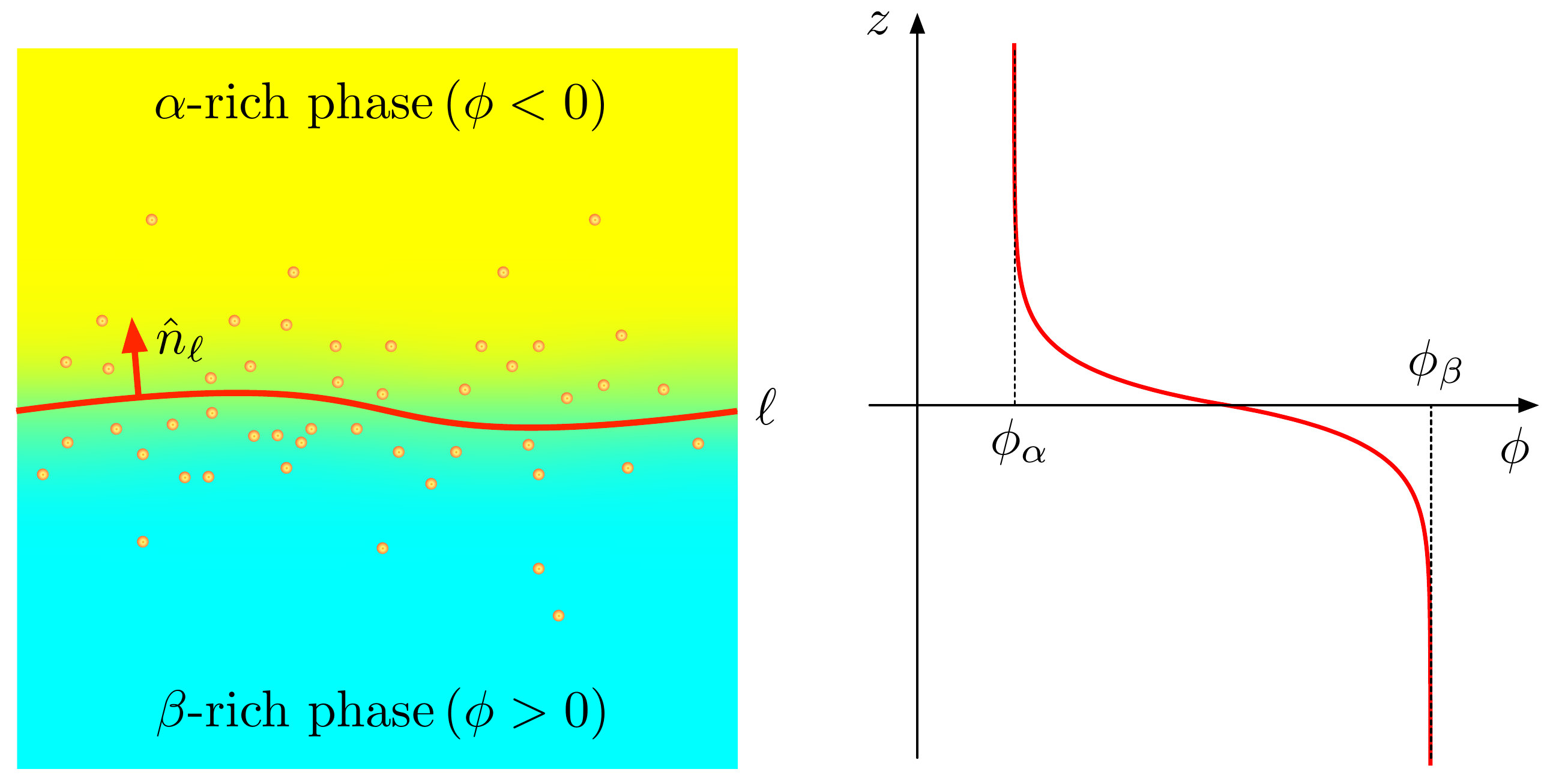}}
\caption{Schematic depiction of the interface $\ell$ separating two coexisting phases (left) and a typical density profile (right).}
\label{fig_interface}
\end{center}
\end{figure}

\section{The model}\label{sec:model}
In this section we outline the construction of the model. The approach we follow consists in two steps: in Sec.~\ref{sec2a} we recall some basic notions from solution theory, hence we consider a model of a homogeneous binary fluid. In Sec.~\ref{sec2b} we extend the model to the inhomogeneous setting and, in parallel, we also model the interaction between the binary fluid and an inhomogeneous solute.

\subsection{Homogeneous mixture}
\label{sec2a}
As a warm up we consider a binary liquid mixture and denote its components with the labels $a$ and $b$. Let $N_{a}$ be the number of molecules of species $a$ and analogously for $N_{b}$. We are assuming incompressibility, hence $N_{a}+N_{b}=N$, with $N$ a constant. 

At constant pressure and temperature, the system is described by the Gibbs free energy $G = G_{0} + \Delta G$, where
\begin{equation}
\begin{aligned}
\label{bin001}
G_{0} & = N_{a}\mu_{a}^{(0)} + N_{b}\mu_{b}^{(0)} \, ,
\end{aligned}
\end{equation}
is the Gibbs free energy of the unmixed system and $\mu_{a}^{(0)}$, and $\mu_{b}^{(0)}$ are the corresponding chemical potentials for pure components $a$ and $b$. The free energy of mixing takes the form \cite{Reichl}
\begin{equation}
\begin{aligned}
\label{bin001-1}
\beta \Delta G & = N_{a}\ln\frac{N_{a}}{N} + N_{b}\ln\frac{N_{b}}{N} + \chi \frac{N_{a}N_{b}}{N} \, ,
\end{aligned}
\end{equation}
where $\beta^{-1}=k_{\textrm{B}}T$ and the (dimensionless) parameter $\chi$ measures the strength of the interaction between the binary mixture components. The chemical potentials $\mu_{j} = \partial G/\partial N_{j}$, with $j=a,b$, are given by
\begin{equation}
\begin{aligned}
\label{bin002}
\mu_{a} & = \mu_{a}^{(0)} + k_{\rm B}T \left( \ln n_{a} + \chi n_{b}^{2} \right) \, , \\
\mu_{b} & = \mu_{b}^{(0)} + k_{\rm B}T \left( \ln n_{b} + \chi n_{a}^{2} \right) \, ,
\end{aligned}
\end{equation}
where $n_{a}=N_{a}/N$, $n_{b}=N_{B}/N$; and evidently, $n_{a}+n_{b}=1$. Parenthetically, we observe that, thanks to Eq.~(\ref{bin002}), the free energy in Eq.~(\ref{bin001}) can be rewritten in the compact form $G = N_{a}\mu_{a}+N_{b}\mu_{b}$. Notice also that the mixing free energy is invariant under the exchange of $N_{a}$ with $N_{b}$. In order to exploit the aforementioned symmetry, we parametrize the concentrations in terms of the order parameter $\phi \in (-1/2,1/2)$, the latter is defined by
\begin{equation}
\begin{aligned}
\label{bin005}
n_{a,b} & = \bar{n} \pm \phi \, , \quad \bar{n}=\frac{1}{2} \, .
\end{aligned}
\end{equation}

At phase coexistence, the two chemical potentials are identical, i.e., $\mu_{a}=\mu_{b}$, hence by defining
\begin{equation}
\label{bin003}
\Delta\mu \equiv \mu_{a}^{(0)} - \mu_{b}^{(0)} \, ,
\end{equation}
from Eq.~(\ref{bin002}) we find the following relationship \footnote{We recall the identity $\tanh^{-1}(x)=\frac{1}{2}\log\frac{1+x}{1-x}$.} between the order parameter and $\Delta\mu$
\begin{equation}
\label{}
\beta\Delta\mu = 2\chi\phi - 2\tanh^{-1}(2\phi) \, .
\end{equation}
The above can be written in the equivalent form
\begin{equation}
\label{bin007}
2\phi = \tanh\bigl[ \chi\phi - \beta \Delta\mu/2 \bigr] \, ,
\end{equation}
the latter is precisely the mean-field equation in the Curie-Weiss model of ferromagnetism, provided $-\Delta\mu$ is identified with the external magnetic field and $\phi$ with the bulk magnetization \cite{Huang}.

The equilibrium composition of the binary mixture follows from the solutions of Eq.~(\ref{bin007}) which corresponds to minima of the Gibbs free energy. On expressing $N_{a}$ and $N_{b}$ in terms of $\phi$, and by expanding in powers of $\phi$, we find
\begin{equation}
\label{bin007_1}
\beta \Delta G/N = 2\phi \tanh^{-1}(2\phi) + \frac{1}{2} \log\Bigl[ \frac{1}{4}-\phi^{2} \Bigr] + \chi \Bigl[ \frac{1}{4}-\phi^{2} \Bigr] \, .
\end{equation}
In order to get a quantitative picture, we expand Eq.~(\ref{bin007_1}) in powers of $\phi$ up to $O(\phi^{4})$, thus Eq.~(\ref{bin007_1}) becomes
\begin{equation}
\begin{aligned}
\label{bin007_2}
\beta \Delta G/N & = W(\phi) + O(\phi^{6}) \, , \\
W(\phi) & = \frac{\chi}{4}-\log2 + (2-\chi) \phi^{2} + \frac{4}{3}\phi^{4} \, ,
\end{aligned}
\end{equation}
thus, we have obtained the standard $\phi^{4}$ double well potential for the bulk free energy. By combining Eq.~(\ref{bin007_2}) and Eq.~(\ref{bin001}), the total Gibbs free energy per particle becomes
\begin{equation}
\begin{aligned}
\label{bin007_3}
\beta G/N & = \left( \bar{n} \beta \mu_{a}^{(0)} + \bar{n} \beta \mu_{b}^{(0)} \right) + W(\phi) + \phi \beta\Delta\mu\, ,
\end{aligned}
\end{equation}
with $\bar{n}$ defined in Eq.~(\ref{bin005}).

Let us consider the phase coexistence, thus we set $\Delta\mu=0$. If $\chi<2$, then the potential Eq.~(\ref{bin007_3}) exhibits a (stable) global minimum only for $\phi=0$, while if $\chi>2$ a new pair of stable solutions of the form $\phi=\pm \phi_{0}$ appears. In the unmixed regime, the equilibrium concentrations are given by $\bar{n} \pm \phi_{0}$ with
\begin{equation}
\label{bin007_a}
\phi_{0} = \sqrt{\frac{3}{8}(\chi-2)} \, .
\end{equation}
The two-phase coexistence line spanned by $\chi>2$ terminates in a critical point, the latter is located in $\chi=2$. The existence of the aforementioned solutions can also be inferred by inspecting the mean-field equation Eq.~(\ref{bin007}) with $\Delta\mu=0$.

\subsection{Inhomogeneous mixture}
\label{sec2b}
In this section we extend the treatment of the binary liquid outlined in Sec. \ref{sec2a} to instances in which inhomogeneous densities are allowed. The typical situation is the one in which the two coexisting phases are separated by an interface. On top of that, we also allow for the presence of third substance, which will be termed the solute.

In order to describe the inhomogeneous system, we promote the number densities to be spatially varying fields, therefore we promote $N_{j}$ to $N_{j}(\textbf{r})$, with $j=a,b$. Then we introduce the densities $\rho_{a}(\textbf{r})=N_{a}(\textbf{r})/V$ and $\rho_{b}(\textbf{r})=N_{b}(\textbf{r})/V$, where $\rho=N/V$ is a constant reference density. The total number of molecules for each species is found by integrating over the volume, i.e., $N_{a}=\int\textrm{d}\textbf{r} \, \rho_{a}(\textbf{r})$, and analogously for $b$.

Having in mind a situation in which the solute is dilute, we model it through an ideal gas with weak interactions with the binary liquid. The Gibbs free energy is now given by
\begin{widetext}
\begin{equation}
\begin{aligned}
\label{bin011}
\beta G[\phi,\varrho] & = \int \textrm{d}\textbf{r} \, \biggl[ \left( \rho_{a}\ln\frac{\rho_{a}}{\rho} + \rho_{b}\ln\frac{\rho_{b}}{\rho} \right) + \chi \frac{\rho_{a}\rho_{b}}{\rho} + \frac{K}{4} (\nabla \rho_{a})^{2} + \frac{K}{4} (\nabla \rho_{b})^{2} + (\omega_{a}\rho_{a}+\omega_{b}\rho_{b}) \varrho \biggr] + \mathcal{H}_{\textrm{gas}}[\varrho;V] \, ,
\end{aligned}
\end{equation}
\end{widetext}
with $\varrho(\textbf{r})$ the solute density and 
\begin{equation}
\label{eq:H_gas}
\mathcal{H}_{\textrm{gas}}[\varrho;V] = \int\textrm{d}\textbf{r} \, \biggl[ \varrho \Bigl( \ln\left( \lambda_{T}^{3}\varrho \right)-1 \Bigr) + \beta V \varrho \biggr] \, ,
\end{equation}
the ideal gas contribution.

Some comments are in order: \emph{i)} although we have introduced two density fields, $\rho_{a}$ and $\rho_{b}$, the functional in Eq.~(\ref{bin011}) depends on a single density field. In fact, $\rho_{a}$ and $\rho_{b}$ are linearly dependent because $\rho_{a}+\rho_{b}=\rho$. The functional given by Eq.~(\ref{bin011}) depends on either $\rho_{a}$, or $\rho_{b}$, or any linear combination of them with the exception of $\rho_{a}+\rho_{b}$. The natural choice is to parametrize the densities by following Eq.~(\ref{bin005}), thus we introduce the order parameter $\phi(\textbf{r})$ defined by
\begin{equation}
\begin{aligned}
\label{bin012}
\rho_{a}(\textbf{r}) & = \frac{\rho}{2}+\rho\phi(\textbf{r}) \, , \\
\rho_{b}(\textbf{r}) & = \frac{\rho}{2}-\rho\phi(\textbf{r}) \, .
\end{aligned}
\end{equation}
\emph{ii)} the coupling $K$ in front of the square gradient terms in Eq.~(\ref{bin011}) quantifies the energetic cost for density inhomogeneities. Within van der Waals theory of liquid-vapor interfaces, the coefficient in front of the square gradient is proportional to the second moment of the inter-particle interaction potential; taking the opposite sign, hence the coefficient is positive \cite{RowlinsonWidom}. Since incompressibility imposes $\rho_{a}(\textbf{r})+\rho_{b}(\textbf{r})=\rho$, the two square gradients in Eq.~(\ref{bin011}) actually coincide.

\emph{iii)} in principle, an external potential $V(\textbf{r})$ can be coupled to the solute density.  The weak interaction between the solute and the binary liquid mixture is taken into account by the coefficients $\omega_{a}$ and $\omega_{b}$. Notice that, if $\omega_{a}=\omega_{b}$, then the solute-binary interaction does not depend on the concentration $\phi$ of the binary liquid, meaning that the solute and the binary liquid are decoupled. On the other hand, if $\omega_{a} \neq \omega_{b}$, then the solute-binary interaction becomes proportional to $\phi(\textbf{r})\varrho(\textbf{r})$. We note that the presence of the potential $V(\textbf{r})$ in Eq.~(\ref{bin011}) makes it different from those models that accounts for a inhomogeneous temperature profile~\cite{Roy2018}.

\emph{iv)} the homogeneous model described in Sec. \ref{sec2a} is retrieved in the double limit $\rho_{a}, \rho_{b} \rightarrow $ constant and $\omega_{a}, \omega_{b} \rightarrow 0$. On the other hand, by turning off the interaction with the solute, and keeping in mind that $\rho_{a}$ and $\rho_{b}$ are not independent, Eq.~(\ref{bin011}) reduces to the Cahn-Hilliard effective Hamiltonian \cite{CH}.

\emph{v}) Since we are interested in inhomogeneous systems, we consider the generalized chemical potentials
\begin{equation}
\label{bin014}
\mu_{j}(\textbf{r}) = \frac{\delta G[\phi,\varrho]}{\delta \rho_{j}(\textbf{r})} \, , \quad j=a,b \, ,
\end{equation}
which are given by \footnote{The occurrence of the Laplacian terms in both $\mu_{a}(\textbf{r})$ and $\mu_{b}(\textbf{r})$ follows from our choice of writing the free energy functional in terms of two densities which, we emphasize, are linearly dependent~\cite{CH}}.
\begin{equation}
\begin{aligned}
\label{bin015}
\beta \mu_{a}(\textbf{r}) & = \ln\frac{\rho_{a}}{\rho} + \chi \frac{\rho_{b}^{2}}{\rho^{2}} - \frac{K}{2} \nabla^{2}\rho_{a} + \omega_{a}\varrho \, , \\
\beta \mu_{b}(\textbf{r}) & = \ln\frac{\rho_{b}}{\rho} + \chi \frac{\rho_{a}^{2}}{\rho^{2}}  - \frac{K}{2} \nabla^{2}\rho_{b} + \omega_{b}\varrho \, .
\end{aligned}
\end{equation}

Again, by taking the limit of homogeneous system with $\omega_{a}=\omega_{b}=0$, we retrieve the results given in Eq.~(\ref{bin002}).

By following the same guidelines of Sec. \ref{sec2a}, in order to find a coarse-grained model, we expand the free energy functional in powers of the density fields $\phi(\textbf{r})$ and $\varrho(\textbf{r})$ by keeping the lowest powers. The result of this procedure leads to the free energy functional
\begin{equation}
\begin{aligned}
\label{bin015_1}
\beta G[\phi,\varrho] & = \rho \int \textrm{d}\textbf{r} \, \biggl[ \frac{K\rho}{2} (\nabla \phi)^{2} + W(\phi) + (\omega_{a}-\omega_{b}) \varrho \phi \biggr] + \\
& + \mathcal{H}_{\textrm{gas}}[\varrho;V_{\varrho}] \, ,
\end{aligned}
\end{equation}
where $V_{\varrho}(\textbf{r})$ is the effective external potential
\begin{equation}
\beta V_{\varrho}(\textbf{r}) = \beta V(\textbf{r}) + \rho\frac{\omega_{a}+\omega_{b}}{2} \, .
\end{equation}
As a consistency check, we observe that for a homogeneous system in absence of the solute, the functional in Eq.~(\ref{bin015_1}) reduces to Eq.~(\ref{bin007_2}), provided $N$ is identified with $\int\textrm{d}\textbf{r} \, \rho$.

The difference of generalized chemical potentials $\Delta\mu_{\phi}(\textbf{r})=\mu_{a}(\textbf{r})-\mu_{b}(\textbf{r})$ follows straightforwardly from Eq.~(\ref{bin015})
\begin{equation}
\begin{aligned}
\label{bin017}
\beta \Delta\mu_{\phi}(\textbf{r}) & = \partial_{\phi}W(\phi)\vert_{\phi(\textbf{r})} - K\rho \nabla^{2} \phi(\textbf{r}) + (\omega_{a}-\omega_{b}) \varrho(\textbf{r}) \, .
\end{aligned}
\end{equation}
Alternatively, Eq.~(\ref{bin017}) follows by taking a functional derivative of the free energy with respect to the order parameter, i.e.,
\begin{equation}
\label{bin018}
\Delta\mu_{\phi}(\textbf{r}) = \frac{\delta G[\phi,\varrho]}{\delta\phi(\textbf{r})} \, .
\end{equation}

It is convenient to factorize the coefficient $K\rho^{2}$ in Eq.~(\ref{bin015_1}), and to introduce the rescaled bulk free energy
\begin{equation}
\label{scaling}
U(\phi) \equiv \frac{W(\phi)}{K\rho} \, ,
\end{equation}
and the rescaled coupling parameter
\begin{equation}
\bar{\omega} \equiv \frac{\omega_{a}-\omega_{b}}{K\rho} \, .
\end{equation}

Thanks to the above definitions, the free energy functional Eq.~(\ref{bin015_1}) can be written as follows
\begin{equation}
\label{001_01}
\beta G[\phi,\varrho] = K\rho^{2} \mathcal{H}[\phi,\varrho] + \mathcal{H}_{\textrm{gas}}[\varrho,V_{\varrho}] \, ,
\end{equation}
with the effective Hamiltonian
\begin{equation}
\label{eff}
\mathcal{H}[\phi,\varrho] = \int \textrm{d}\textbf{r} \biggl[ \frac{1}{2} (\nabla\phi(\textbf{r}))^{2} + U(\phi(\textbf{r})) + \bar{\omega} \phi(\textbf{r})\varrho(\textbf{r}) \biggr] \, .
\end{equation}
Notice that Eq.~(\ref{eff}) is formally identical to the Landau-Ginzburg free energy for a uniaxial ferromagnet with order parameter field $\phi(\mathbf{r})$ in the presence of an inhomogeneous bulk field $h(\textbf{r}) = - \bar{\omega} \varrho(\textbf{r})$ \cite{PRBRE2006, BPRRE2009}. 

\subsection{Equilibrium}
\label{sec2c}
At thermodynamic equilibrium, the chemical potentials of both components are equal, therefore their difference given by Eq.~(\ref{bin018}) vanishes. Moreover, the chemical potential of the solute
\begin{equation}
\begin{aligned}
\label{}
\mu_{\varrho}(\textbf{r}) & = \frac{\delta G[\phi,\varrho]}{\delta\varrho(\textbf{r})} \, ,
\end{aligned}
\end{equation}
is constant. A simple calculation gives
\begin{equation}
\beta\mu_{\varrho} = \log(\lambda_{T}^{3}\varrho(\textbf{r})) + \beta V(\textbf{r}) + \omega_{a}\rho_{a}(\textbf{r}) + \omega_{b}\rho_{b}(\textbf{r}) \, ,
\end{equation}
and therefore the solute density is given by a Boltzmann distribution, which we write in the compact form
\begin{equation}
\label{05bis}
\varrho(\textbf{r}) = \varrho_{0} \textrm{e}^{-\beta \mathcal{V}(\textbf{r})} \, ,
\end{equation}
where
\begin{equation}
\varrho_{0} = \lambda_{T}^{-3} \textrm{e}^{\beta\mu_{\varrho}}
\end{equation}
is a reference density set by the chemical potential $\mu_{\varrho}$, while
\begin{equation}
\label{bin018_001}
\beta\mathcal{V}(\textbf{r}) = \beta V(\textbf{r}) + \omega_{a}\rho_{a}(\textbf{r}) + \omega_{b}\rho_{b}(\textbf{r})
\end{equation}
is the total external potential acting on the solute. Notice that $\mathcal{V}(\textbf{r})$ receives a contribution also from the binary liquid provided $\omega_{a,b}\neq0$.

Concerning the density of the binary mixture, the latter turns out to be found as the solution of
\begin{equation}
\label{}
\Delta\mu_{\phi}(\textbf{r}) = 0 \, ,
\end{equation}
which follows by equating the two generalized chemical potentials in Eq.~(\ref{bin015}). Thanks to Eq.~(\ref{bin017}), the above gives
\begin{equation}
\label{bin019}
K\rho \nabla^{2} \phi(\textbf{r}) - \partial_{\phi}W(\phi)\vert_{\phi(\textbf{r})} = (\omega_{a}-\omega_{b}) \varrho(\textbf{r}) \, .
\end{equation}
We observe that a nonzero solute density $\varrho(\textbf{r})$ acts as an external source for the binary liquid density $\phi(\textbf{r})$. As a result, the binary liquid density changes accordingly due to the presence of the solute. The solution of Eq.~(\ref{bin019}) will be discussed in Sec.~\ref{sec3}. We conclude by noting that in the absence of the solute-binary interaction, i.e., $\omega_{a}=\omega_{b}$, the right hand side of Eq.~(\ref{bin019}) vanishes and the density profile of the binary liquid is given by the classical van der Waals theory \cite{RowlinsonWidom}. 

\subsection{Out of equilibrium}
Let us consider now a \emph{non}-equilibrium regime in which the solute is produced within the binary mixture, while the latter is assumed to be at local equilibrium for the assigned solute density $\varrho(\textbf{r})$. Within this scenario, the production rate of the solute is considered to be rather slow, thus the binary liquid adjusts its composition by following the solute density according to Eq.~(\ref{bin019}). Accordingly, the binary mixture is at equilibrium with the specified solute density profile and hence there are no density currents within the binary mixture, i.e., $\textbf{J}_{\phi} \propto \nabla \Delta\mu_{\phi} = \textbf{0}$, the latter follows from $\Delta\mu_{\phi}=0$. The density $\phi(\textbf{r})$ satisfies Eq.~(\ref{bin019}), however the source term $\propto \varrho(\textbf{r})$ has to be identified, as we are going to show.

Contrary to the equilibrium setting discussed in Sec. \ref{sec2c}, the chemical potential of the solute is no longer constant, therefore it exists the net current of solute
\begin{equation}
\label{08}
\textbf{J}_{\varrho}(\textbf{r}) = - L_{\varrho} \nabla \beta\mu_{\varrho}(\textbf{r}) \, ;
\end{equation}
within linear response theory, the Onsager coefficient reads $L_{\varrho}=D\varrho$ where $D$ is a diffusion coefficient, thus Eq.~(\ref{08}) becomes
\begin{equation}
\label{09}
\textbf{J}_{\varrho}(\textbf{r}) = - D \nabla\varrho(\textbf{r}) - \beta D \varrho(\textbf{r}) \nabla \mathcal{V}(\textbf{r}) \, .
\end{equation}

The time evolution of the solute density then follows accordingly with the continuity equation if no reactions are taking place, and if the solute is not produced/destroyed within the system, i.e., $\partial_{t}\varrho = - \nabla\cdot\textbf{J}_{\varrho}$. More generally, however, the solute is produced/destroyed due to certain chemical reactions whose occurrence is confined within small region of the system; hence sources and sinks have to be included in the description. We thus consider the following equation for the time evolution of the solute density
\begin{equation}
\label{09_1}
\partial_{t}\varrho(\textbf{r}) + \nabla\cdot\textbf{J}_{\varrho}(\textbf{r}) = S(\textbf{r}) - \varrho(\textbf{r})/\tau \, .
\end{equation}
The terms on the right hand side of Eq.~(\ref{09_1}) correspond to the density of sources, $S(\textbf{r})$, and sinks, $\varrho(\textbf{r})/\tau$, where we have assumed a uniform decay of the solute concentration with characteristic time $\tau$. It is understood that $S(\textbf{r})$ may take both signs, thus a local sink would correspond to $S(\textbf{r})<0$ in certain regions of space. We will see an example of this sort in Sec. \ref{3b}.

By combining Eq.~(\ref{09}) and Eq.~(\ref{09_1}), the non-equilibrium stationary state
is given by the solution of
\begin{equation}
\label{09_2}
D\nabla^{2}\varrho(\textbf{r}) +\beta D \nabla \cdot \Bigl[ \varrho(\textbf{r}) \nabla \mathcal{V}(\textbf{r}) \Bigr] = \varrho(\textbf{r})/\tau - S(\textbf{r}) \, ,
\end{equation}
with asymptotic boundary condition $\varrho \rightarrow0$ at $|\textbf{r}|\rightarrow\infty$ (away from sources/sinks) and with a specific boundary condition on the sources/sinks.

We emphasize that Eq.~(\ref{09_2}) and Eq.~(\ref{bin019}) form a coupled system of differential equations for the densities $\phi(\textbf{r})$ and $\varrho(\textbf{r})$. In particular, from Eq.~(\ref{bin019}) we see how the density $\phi(\textbf{r})$ decouples from $\varrho(\textbf{r})$ provided $\omega_{a}=\omega_{b}$. Analogously, if $\omega_{a}=\omega_{b}$, then the total potential $\mathcal{V}(\textbf{r})$ does not depend on $\phi(\textbf{r})$ and, correspondingly, also Eq.~(\ref{09_2}) decouples from $\phi(\textbf{r})$.

The above considerations suggest to seek for a solution for the density fields $\phi(\textbf{r})$ and $\varrho(\textbf{r})$ by means of a perturbative analysis based on the parameter $\bar{\omega}$, which is assumed to be small. In order to illustrate the approach, we consider the absence of an external potential, i.e., $V(\textbf{r})=0$, and write the solution of Eq.~(\ref{09_2}) as follows
\begin{equation}
\varrho(\textbf{r}) = \varrho_{0}(\textbf{r}) + \varrho_{1}(\textbf{r}) + \varrho_{2}(\textbf{r}) + \dots \, ,
\end{equation}
in which $\varrho_{k}(\textbf{r})=O(\bar{\omega}^{k})$. At leading-order in powers of $\bar{\omega}$, Eq.~(\ref{09_2}) reads
\begin{equation}
\label{42}
\left( - D\nabla^{2} + \tau^{-1} \right)\varrho_{0}(\textbf{r}) = S(\textbf{r}) \, .
\end{equation}
The above will be solved in Sec. \ref{3b} for a specific choice of the solute source $S(\textbf{r})$.

\section{Density profiles in the presence of an external field}
\label{sec3}
\subsection{General formalism}
In this section we analyze the binary liquid mixture in the phase separated regime in the presence of an external field which, within our formalism, is represented by the density of solute. In particular, we show how to compute the density $\phi$ across a domain wall separating two coexisting phases. We employ the so-called double parabola approximation, which consists in replacing the $\phi^{4}$ double well with two parabolic branches, i.e.,
\begin{equation}
\label{potentialdp}
U(\phi) = \frac{\kappa^{2}}{2} \left( |\phi|-\phi_{0} \right)^{2} \, .
\end{equation}
The latter has been successfully employed in the study of short range critical wetting in three dimensions \cite{PRBRE2006, PRBRE2007,PRBRE2008}, for which a wealth of formal results can be applied to the model studied in this paper.

To be definite, we consider a phase-separated binary mixture with the component of higher density standing below the interface, as depicted in Fig.\ref{fig_interface}. The mean-field equation governing the density profile in the presence of the external field can be found by imposing the stationarity of the functional given by Eq.~(\ref{eff})
\begin{equation}
\label{C_01}
\frac{\delta \mathcal{H}[\phi,\varrho] }{\delta \phi(\textbf{r})} \bigg\vert_{\phi_{\Xi}(\textbf{r})} = 0 \, ,
\end{equation}
in which $\phi_{\Xi}(\textbf{r})$ denotes the solution of Eq.~(\ref{C_01}) with the appropriate boundary conditions. The advantage of the double-parabola approximation is that it allows for an analytical treatment, because Eq.~(\ref{eff}) gives a pair of linear partial differential equations, which can be solved by means of Green's functions techniques. Eq.~(\ref{C_01}) gives
\begin{equation}
\label{C_03}
\begin{cases}
(-\nabla^{2} + \kappa^{2})(\phi_{\Xi}(\textbf{r})-\phi_{0})=-\bar{\omega} \varrho(\textbf{r}) \, , & \textbf{r} \in \Omega_{a} \, , \\
(-\nabla^{2} + \kappa^{2})(\phi_{\Xi}(\textbf{r})+\phi_{0})=-\bar{\omega} \varrho(\textbf{r}) \, , & \textbf{r} \in \Omega_{b} \, ,
\end{cases}
\end{equation}
where $\Omega_{a}$ denotes the region occupied by the $a$-rich phase, i.e. $\phi>0$, while $\Omega_{b}$ denotes the region occupied by the $b$-rich phase, i.e. $\phi<0$. The above differential equations have to be supplemented with the asymptotic boundary conditions $\phi(\textbf{r}) \rightarrow \pm \phi_{0}$ for $z \rightarrow \mp \infty$. Another pair of boundary conditions follows by imposing $\phi(\textbf{r}_{\ell})=0$ along the interface, the latter corresponds to the locus of points in which the domains $\Omega_{a}$ and $\Omega_{b}$ come in touch. According to the crossing criterion mentioned above, the interface is formally defined by
\begin{equation}
\label{C_04}
\ell \coloneqq \{ \textbf{r}_{\ell} \in \mathbb{R}^{3} \, \vert \, \phi(\textbf{r}_{\ell})=0 \} \, .
\end{equation}
We specialize our attention to a flat interface parallel to the $(x,y)$ plane, and let $z=\ell$ be the interface location. The regions $\Omega_{a}$ and $\Omega_{b}$ are respectively given by the half-spaces $z<\ell$ and $z>\ell$, while the plane $z=\ell$ defines the locus of the interface.

In order to solve the non-homogeneous problem specified by Eq.~(\ref{C_03}), we first consider the homogeneous one ($\varrho=0$). The rescaled Green function $K(\textbf{r},\textbf{r}^{\prime})$ satisfies
\begin{equation}
\label{C_07}
\left( - \nabla_{\textbf{r}}^{2} + \kappa^{2} \right) K(\textbf{r},\textbf{r}^{\prime}) = 2\kappa \delta(\textbf{r}-\textbf{r}^{\prime}) \, ,
\end{equation}
where $\kappa=\xi_{\textrm{b}}^{-1}$ is the inverse bulk correlation length. In three spatial dimensions, the above gives the Ornstein-Zernike kernel \cite{PRBRE2006}
\begin{equation}
\label{C_08}
K(\textbf{r},\textbf{r}^{\prime}) = \frac{ \kappa \exp(-\kappa |\textbf{r}-\textbf{r}^{\prime}|) }{ 2\pi |\textbf{r}-\textbf{r}^{\prime}| } \, .
\end{equation}
By virtue of translational invariance along the $(x,y)$ directions, the three-dimensional problem can be reduced to a one-dimensional problem. By integrating Eq.~(\ref{C_08}) over the longitudinal coordinates ($x$ and $y$), we obtain the rescaled Green's function
\begin{equation}
\label{C_09}
\mathcal{K}(z, z^{\prime}) = \exp{\left( -\kappa |z-z^{\prime}| \right)} \, ,
\end{equation}
which satisfies the one-dimensional version of Eq.~(\ref{C_07})
\begin{equation}
\label{C_15}
\left( -\partial_{z}^{2}+\kappa^{2} \right) \mathcal{K}(z,z^{\prime}) = 2\kappa\delta(z-z^{\prime}) \, .
\end{equation}

\begin{figure}[htbp]
\begin{center}
\centerline{\includegraphics[width = .46 \textwidth]{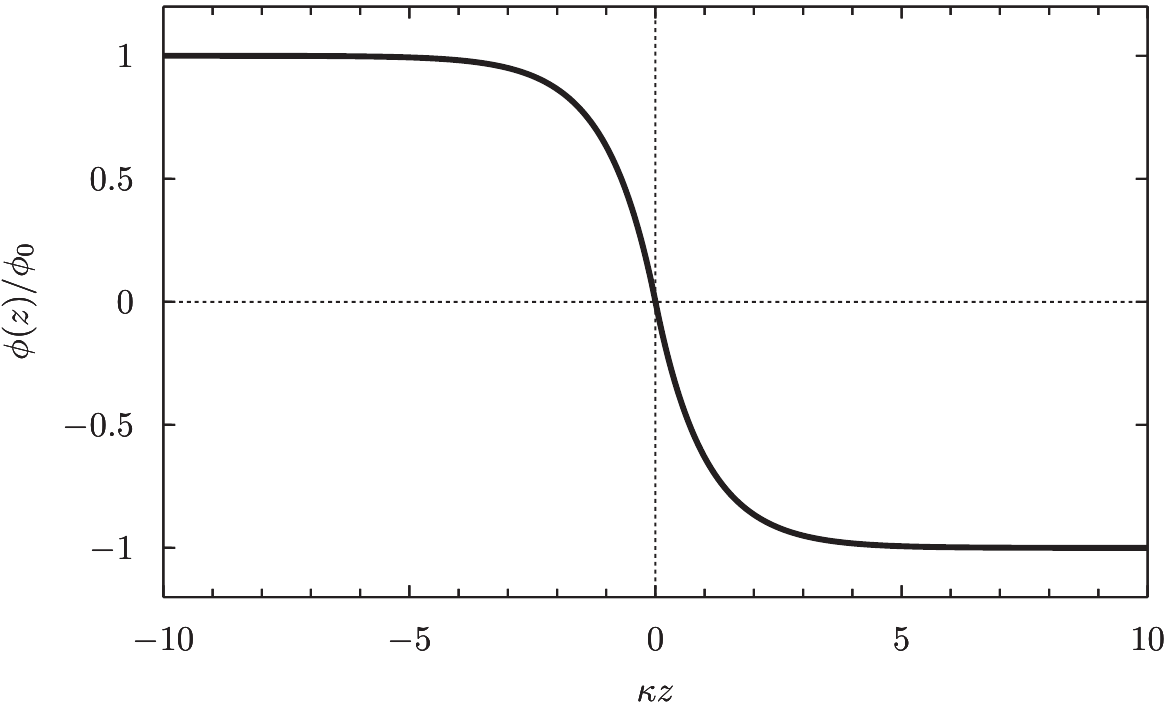}}
\caption{The interface profiles Eq.~(\ref{profile_DP}) corresponding to a domain wall in the binary mixture in the absence of a solute.}
\label{fig_freeinterface}
\end{center}
\end{figure}

The solution of Eq.~(\ref{C_03}) with asymptotic conditions $\phi(z \rightarrow \pm \infty) = \mp \phi_{0}$ enforcing a domain wall, can be split as follows
\begin{equation}
\label{C_10}
\phi_{\Xi}(z;\varrho) = \phi_{\Xi}(z;0) + \Delta\phi_{\Xi}(z;\varrho) \, ,
\end{equation}
where $\phi_{\Xi}(z;0)$ is the density profile for the homogeneous problem, i.e., when $\varrho=0$, which reads
\begin{equation}
\label{profile_DP}
\phi_{\Xi}(z;0) = 
\begin{cases}
\phi_{0} - \phi_{0} \textrm{e}^{-\kappa(\ell-z)} \, , & z<\ell, \\
\\
-\phi_{0} + \phi_{0} \textrm{e}^{-\kappa(z-\ell)} \, , & z>\ell \, ,
\end{cases}
\end{equation}
where $\ell$ defines the interface position. The density profile given by Eq.~(\ref{profile_DP}) is plotted in Fig. \ref{fig_freeinterface}.

The interface profile given in Eq.~\eqref{profile_DP} smoothly interpolates between the bulk phases $\pm \phi_{0}$. The typical variations in density occur on the length scale set by the bulk correlation length. We also point out that $\ell$ is not fixed for an isolated interface standing in the bulk; this is a natural consequence of invariance under translations along the $z$ axis. The archetypal situation in which the translational invariance is broken is provided by the presence of a wall. Such an instance occurs within the study of phase separation in confined geometries and wetting transitions. There the interface height $z=\ell$ with respect to a wall located in $z=0$ turns out to be determined by the adsorption properties of the wall; see for instance \cite{Dietrich}. In our problem instead, $\ell$ will be determined by a free energy minimization compatible with a canonical criterion in which the mass of the binary liquid is conserved (see Sec. \ref{3b}).

Coming back to Eq.~(\ref{C_10}), the term $\Delta\phi_{\Xi}(z;\varrho)$ gives the deviation from the homogeneous solution. The latter satisfies the inhomogeneous problem
\begin{equation}
\begin{aligned}
\label{C_16}
\left( -\partial_{z}^{2}+\kappa^{2} \right) \Delta\phi_{\Xi}(z;\varrho) & = -  \bar{\omega}  \varrho(z) \, ,
\end{aligned}
\end{equation}
with boundary condition
\begin{equation}
\label{C_12}
\Delta\phi_{\Xi}(\ell;\varrho) = 0
\end{equation}
implied by the crossing criterion ($\phi_{\Xi}(\ell;\varrho) = 0$) together with the decomposition Eq.~(\ref{C_10}). The solution of Eq.~(\ref{C_16}) subjected to Eq.~(\ref{C_12}), is given by
\begin{widetext}
\begin{equation}
\label{C_13}
\Delta\phi_{\Xi}(z;\varrho) = -\frac{  \bar{\omega}  }{ 2 \kappa }
\begin{cases}
\int_{-\infty}^{\ell} \textrm{d}z^{\prime} \, \varrho(z^{\prime}) \mathcal{K}(z^{\prime},z) - \int_{-\infty}^{\ell} \textrm{d}z^{\prime} \, \varrho(z^{\prime}) \mathcal{K}(z^{\prime},\ell) \mathcal{K}(\ell,z) \, ,  & z<\ell \\
\\
\int_{\ell}^{\infty} \textrm{d}z^{\prime} \, \varrho(z^{\prime}) \mathcal{K}(z^{\prime},z) - \int_{\ell}^{\infty} \textrm{d}z^{\prime} \, \varrho(z^{\prime}) \mathcal{K}(z^{\prime},\ell) \mathcal{K}(\ell,z) \, , & z>\ell \, .
\end{cases}
\end{equation}
\end{widetext}
In order to prove that Eq.~(\ref{C_13}) is the correct solution, it is enough to apply the operator $-\partial_{z}^{2}+\kappa^{2}$ by using Eq.~(\ref{C_15}) and observe that the term with two $\mathcal{K}$-kernels vanishes because $z \neq \ell$. For $z>\ell$ the second square bracket vanishes, while the first one gives $\varrho(z)$. A similar reasoning applies to the case $z<\ell$, now with the first square bracket vanishes while the second one gives $\varrho(z)$.

\subsection{Density profiles due to a source-sink doublet}
\label{3b}
We are now in the position to apply the ideas developed so far. The interface hosts chemical reaction with the solute which, in turn, is produced on one side of the interface and then it disappears on the other. This specific instance is modeled by introducing sources of solute on one side of the interface and sinks on the other one; hence a source-sink doublet.

In order to single out the essential features of the model, we consider a flat interface. Consequently, any density inhomogeneity will depend only on $z$. By virtue of translational invariance along the directions $x$ and $y$, Eq.~(\ref{42}) gives
\begin{equation}
\left( - D\partial_{z}^{2} + \tau^{-1} \right)\varrho_{0}(z) = S(z) \, ,
\end{equation}
with the source-sink doublet modeled via
\begin{equation}
\label{doublet}
S(z)=S_{0}\delta(z-a)-S_{0}\delta(z+a) \, .
\end{equation}
The magnitude $S_{0}>0$ corresponds to a source localized in $z=a$ and a sink localized in $z=-a$. It is clear that a simultaneous sign reversal of both $S_{0}$ and $a$ is a symmetry of Eq.~(\ref{doublet}). The specific functional form of localized source/sinks given by (\ref{doublet}) has been used for mathematical convenience. It is indeed simple to extend the calculation to the case in which Eq.~(\ref{doublet}) is replaced by a smooth function, for instance given by a Gaussian profile centered in $z=a$ and another one with opposite amplitude centered in $z=-a$.

By introducing the Fourier transform
\begin{equation}
\widehat{\varrho}_{0}(k) = \int_{-\infty}^{+\infty} \textrm{d}z \, \varrho_{0}(k) \textrm{e}^{-ikz} \, ,
\end{equation}
with inverse
\begin{equation}
\label{FT01}
\varrho_{0}(z) = \int_{-\infty}^{+\infty} \frac{\textrm{d}k}{2\pi} \widehat{\varrho}_{0}(k) \textrm{e}^{ikz} \, ,
\end{equation}
and analogously for $S(z)$, the density of solute in the stationary state is formally given by
\begin{equation}
\label{FT02}
\varrho_{0}(z) = \int_{-\infty}^{+\infty} \frac{\textrm{d}k}{2\pi} \frac{ \widehat{S}(k) }{ \tau^{-1} + Dk^{2} } \textrm{e}^{ikz} \, .
\end{equation}
The Fourier transform of Eq.~(\ref{doublet}) follows straightforwardly
\begin{equation}
\label{FT04}
\widehat{S}(k) = - 2i S_{0} \sin(ka) \, ,
\end{equation}
therefore Eq.~(\ref{FT02}) immediately leads us to
\begin{equation}
\label{FT05}
\varrho(z) = \bar{\varrho}_{0} \left( \textrm{e}^{-|z-a|/\lambda} - \textrm{e}^{-|z+a|/\lambda} \right) \, ,
\end{equation}
with penetration depth
\begin{equation}
\label{depthscale}
\lambda=\sqrt{D\tau}
\end{equation}
and overall amplitude $\bar{\varrho}_{0}=S_{0}\lambda/(2D)$.

Let us consider the interface separating the components of the binary mixture and the density profile across it. At order $O(\bar{\omega}^{0})$ the density profile is the one corresponding to a free interface, the latter is given by $\phi_{\Xi}(z;0)$; see Fig. \ref{fig_freeinterface}. At order $O(\bar{\omega})$ the density profile is given by
\begin{equation}
\label{n100}
\phi_{\Xi}(z;\varrho) = \phi_{\Xi}(z;0) + \Delta\phi(x;\varrho_{0}) + O(\bar{\omega}^{2}) \, ,
\end{equation}
notice that in the above we have plugged the leading order solution $\varrho_{0}$ in the correction term $\Delta\phi(x;\varrho)$, thus that contribution is of order $O(\bar{\omega})$. By plugging the solute profile given by Eq.~(\ref{FT05}) into Eq.~(\ref{n100}), we obtain a family of profiles parametrized by the interface location $\ell$. The position of the interface $\ell$ is obtained by imposing the mass conservation of the two species. In our model this implies that
\begin{equation}
\label{def201}
M=\lim_{L\rightarrow\infty} \int_{-L}^{L}\textrm{d}z \, \phi(z;\varrho) = 0 \, .
\end{equation}
Such a constraint is satisfied in the absence of solute by the free profile $\phi_{\Xi}(z;0)$ by taking $\ell=0$.

\begin{figure}[htbp]
\begin{center}
\centerline{\includegraphics[width = .5 \textwidth]{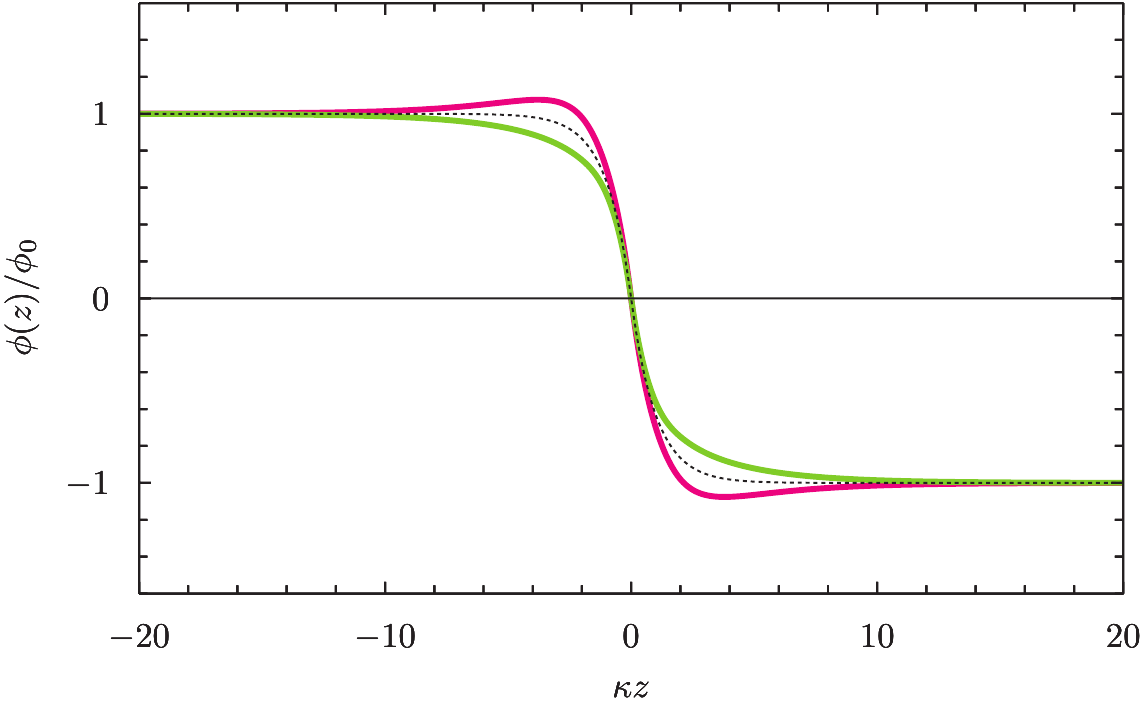}}
\caption{Density profile for the source/sink doublet centered in $z=0$ with source/sink separation $\kappa a=2$, penetration length $\kappa\lambda=3$ and strength $\bar{\omega}\bar{\varrho}_{0}/(\kappa\phi_{0}^{2}) = \pm 0.25$ (see Eq. (\ref{eq:def-g})), respectively in pink and green. The dotted curve corresponds to the unperturbed profile, the one given by Eq.~(\ref{profile_DP}).}
\label{fig_profiles2}
\end{center}
\end{figure}

Setting $\ell=0$ in the presence of the solute still gives a vanishing total mass because the resulting profile is odd with respect to $z=0$. Therefore the interface of the binary liquid exhibits the profiles shown in Fig. \ref{fig_profiles2}.

The density profiles of Fig.~\ref{fig_profiles2} exhibit significant deviations from the unperturbed profile of Fig.~\ref{fig_freeinterface} only within a finite distance within the interfacial region. The above feature is actually a direct consequence of the fast decay which characterizes the solute density $\varrho(z)$; see Eq.~\eqref{FT05}. Notice also how the binary-solute interaction can lead to both monotonic ($\omega_{a}<\omega_{b}$) and non monotonic ($\omega_{a}>\omega_{b}$) density profiles.

\section{Free energy in the presence of an external field}\label{sec4}
In this section we examine the energetic aspects by computing the free energy for the binary liquid in the presence of the solute. It is instructive to begin by considering the case in which the solute is absent \cite{RowlinsonWidom}. Then we will introduce a non vanishing solute concentration and we will examine its consequences on the free energy.

\subsection{Surface tension in the presence of an inhomogeneous field}
\label{freeinterface}
In the absence of external fields, the surface tension is defined as the difference - per unit area - between the free energy of the system with an interface and the free energy of the system without the interface. The excess free energy (in $k_{\rm B}T$ units) is thus given by
\begin{equation}
\label{efe2}
\delta \mathcal{F} = \beta G[\phi_{\Xi};0] - \beta G[\phi_{0};0] \, ,
\end{equation}
where $G[\phi_{0};0]$ is the free energy of the system characterized by the uniform density $\phi_{0}$. For a flat interface parallel to the $xy$-plane we can perform the integrations along $x$ and $y$, this gives the cross sectional area $\mathcal{A}_{\pi} = \int\textrm{d}x\textrm{d}y$, which factorizes. Hence the surface tension is given by
\begin{equation}
\label{15}
\gamma_{0} = \frac{ \delta\mathcal{F} }{ \mathcal{A}_{\pi} } \, .
\end{equation}
The calculation of Eq.~\eqref{15} is a standard exercise which we briefly recall. Equation (\ref{15}) is actually the functional
\begin{equation}
\label{16}
\gamma_{0} = K\rho^{2} \int_{-\infty}^{+\infty}\textrm{d}z \biggl[ \frac{1}{2} \left( \frac{\textrm{d}\phi_{\Xi}}{\textrm{d}z} \right)^{2} + U(\phi_{\Xi}) \biggr] \, ,
\end{equation}
in which $\phi_{\Xi}$ is the profile Eq.~\eqref{profile_DP}. It is well known that for ``one-dimensional'' interfaces the above equation admits a mechanical interpretation in which the energy minimization is equivalent to Hamilton's principle of least action \cite{RowlinsonWidom}. From this mechanical analogy it follows that Eq.~(\ref{16}) admits a first integral. As a result, the contribution stemming from the square gradient - the kinetic term in the mechanical analogy - and the contribution due to the potential in Eq.~(\ref{16}), actually coincide. Moreover, one can show that
\begin{equation}
\label{18}
\gamma_{0} = K\rho^{2} \int_{-\phi_{0}}^{\phi_{0}}\textrm{d}\phi \sqrt{2U(\phi)} \, ,
\end{equation}
which implies that $U(\phi)$ suffices for the determination of the surface tension. By using Eq.~(\ref{18}) it is thus immediate to obtain
\begin{equation}
\label{20}
\gamma_{0} = K\rho^{2} \kappa \phi_{0}^{2} \, .
\end{equation}
If we would have used the $\phi^{4}$ double well we would have obtained the same result up to a factor $2/3$.

It is worth notice that for an effective Hamiltonian which is invariant under the global transformation $\phi \leftrightarrow -\phi$, then the excess free energy
\begin{equation}
\label{efe3prime}
\delta^{\prime} \mathcal{F}_{\Xi} = \beta G[\phi_{\Xi};0] - \beta G[\phi_{\rm sharp};0] \, ,
\end{equation}
with
\begin{equation}
\label{efe3}
\phi_{\textrm{sharp}}(z;\ell)
=
\begin{cases}
\phi_{0}      & z<\ell \, , \\
-\phi_{0}      & z>\ell \, ,
\end{cases}
\end{equation}
coincides with the one appearing in Eq.~(\ref{efe2}) if we neglect a set of zero measure in which the gradient is ill defined. The term subtracted in Eq.~(\ref{efe3prime}) corresponds to the bulk free energy of a uniform system with density $\phi_{0}$ for $z<\ell$ and density $-\phi_{0}$ for $z>\ell$. Since that $\delta \mathcal{F}_{\Xi}=\delta^{\prime} \mathcal{F}_{\Xi}$, both Eq.~\eqref{efe2} and Eq.~\eqref{efe3prime} lead to the same value of the surface tension.

In order to construct the notion (which is not unique) of excess free energy in the presence of an external field, we adapt the definition Eq.~\eqref{efe3prime} as follows
\begin{equation}
\label{efe4prime}
\delta^{\prime} \mathcal{F} = \beta G[\phi_{\Xi};\varrho] - \beta G[\phi_{\rm sharp};\varrho] \, ,
\end{equation}
thus the excess free energy per unit area (per $k_{\rm B}T$) reads
\begin{equation}
\label{efe4prim1}
\gamma[\varrho] = \frac{ \delta^{\prime} \mathcal{F} }{ \mathcal{A}_{\pi} } \, .
\end{equation}
Taking the limit $\varrho \rightarrow 0$ we have $\gamma[\varrho] \rightarrow \gamma_{0}$, thus we retrieve the result for unperturbed interfaces, as it should be. The excess free energy Eq.~\eqref{efe4prime} can be formulated in terms of the effective Hamiltonian Eq.~\eqref{eff}. Thanks to the results summarized in Appendix~\ref{appendix}, up to terms of order $\bar{\omega}$, Eq.~\eqref{efe4prim1} becomes
\begin{equation}
\label{def203}
\gamma[\varrho] = \gamma_{0} + \gamma_{1}[\varrho] \, ,
\end{equation}
with
\begin{equation}
\begin{aligned}
\label{newgamma1}
\gamma_{1}[\varrho] & = - K\rho^{2} \bar{\omega} \phi_{0} \int_{-\infty}^{\ell}\textrm{d}z \, \varrho_{0}(z) \mathcal{K}(z,\ell) \\
& + K\rho^{2} \bar{\omega} \phi_{0} \int_{\ell}^{\infty}\textrm{d}z \, \varrho_{0}(z) \mathcal{K}(z,\ell) \, .
\end{aligned}
\end{equation}
We observe that $\mathcal{K}(z,\ell)=\exp(-|z-\ell|/\xi_{\rm b})$ vanishes exponentially far away from the interface position. It thus follows that both terms in (\ref{A08}) are always finite since the solute density is expected to be bounded far away from $z=\ell$.

\subsection{Modified surface tension due to a source-sink doublet}
\begin{figure}[htbp]
\begin{center}
\centerline{\includegraphics[width = .5 \textwidth]{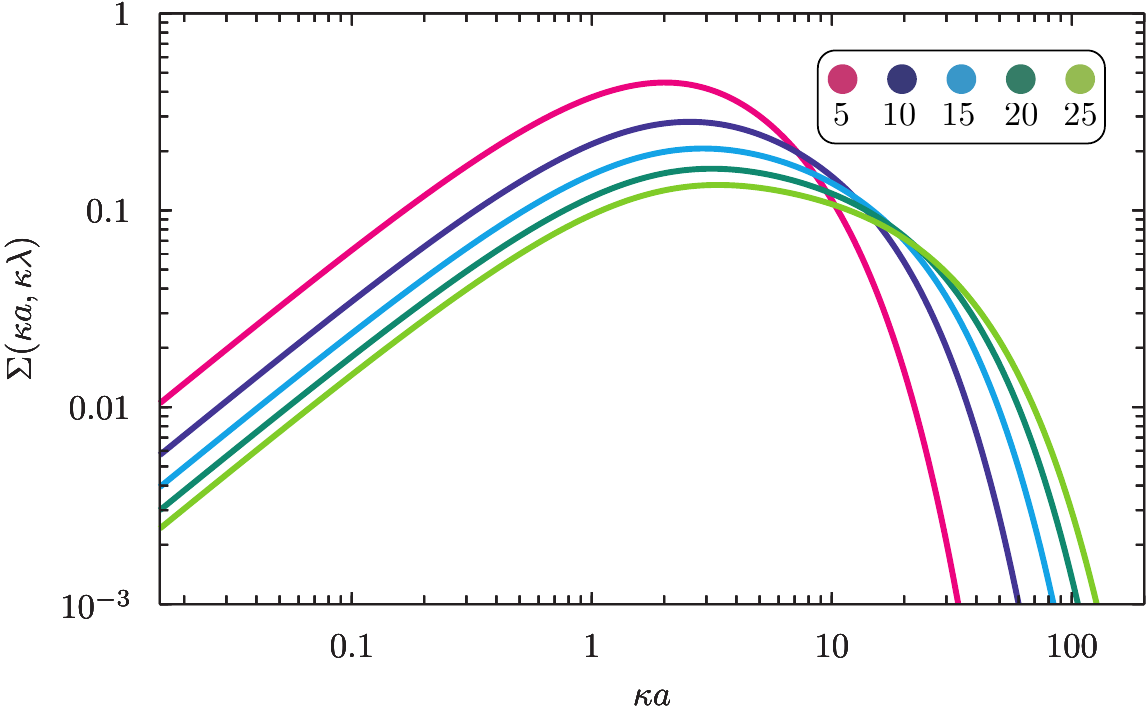}}
\caption{The positive branch $(a>0)$ of the scaling function $\Sigma(\kappa a,\kappa \lambda)$. The curves correspond to different values of the rescaled penetration depth $\kappa \lambda=5,10,15,20,25$ as indicated in the legend.}
\label{fig_logtension2}
\end{center}
\end{figure}

As a concrete example, we compute the deviation from the unperturbed surface tension $\gamma_{1}[\varrho]$ for the source-sink doublet we discussed in Sec. \ref{3b}. It is convenient to write $\gamma_{1}[\varrho]$ in units of $\gamma_{0}$ and express the result as follows
\begin{equation}
\gamma_{1}[\varrho]/\gamma_{0} = g \Sigma(\kappa a,\kappa\lambda) \, ,
\end{equation}
where $g$ is the dimensionless parameter
\begin{equation}
\label{eq:def-g}
g=\bar{\omega}\bar{\varrho}_{0}/(\kappa^{2}\phi_{0})
\end{equation}
which measures the strength of the solute-binary liquid coupling, while $\Sigma(\kappa a,\kappa\lambda)$ is a dimensionless scaling function of the scaling variables $\kappa\lambda=\lambda/\xi_{\rm b}$ and $\kappa a=a/\xi_{\rm b}$; we recall that $\xi_{\rm b}$ is the bulk correlation length. A simple calculation gives
\begin{equation}
\label{scalingdouble}
\Sigma(\kappa a,\kappa\lambda) = \frac{4\kappa\lambda}{\kappa^{2}\lambda^{2}-1}\left( \textrm{e}^{-|a|/\lambda} - \textrm{e}^{-\kappa|a|} \right) \textrm{sign}(a) \, .
\end{equation}
Notice that Eq.~(\ref{scalingdouble}) is an odd function of the distance, $a$, of the source/sink from the interface. The scaling function Eq.~(\ref{scalingdouble}) is plotted in Fig. \ref{fig_logtension2}.

For a fixed penetration depth $\lambda$ and correlation length $\kappa$ (see Eqs. \eqref{FT05} and \eqref{depthscale}), the modified surface tension $\gamma_{1}[\varrho]$ exhibits a maximum when the separation between sources and sinks attains the value
\begin{equation}
\label{eq:def-k}
\kappa a_{\max} = \frac{\kappa\lambda}{\kappa\lambda-1}\ln(\kappa\lambda) \, .
\end{equation}
The latter gives rise to the maximum (for $g>0$)
\begin{equation}
\gamma^\text{(max)}_{1}[\varrho] / \gamma_{0} = \frac{4g}{1+\kappa\lambda} (\kappa\lambda)^{1/(1-\kappa\lambda)} \, .
\end{equation}
Regarding the above as a function of $\kappa\lambda$, it is simple to show that it reaches the maximum value $8g/\textrm{e} \approx 2.9 g$ for $\kappa\lambda=1$. The maximum deviation of the surface tension is thus reached when the penetration depth matches the bulk correlation length. The explicit solution shown above indicates that $\gamma_{1}[\varrho]$ becomes a strong effect only within a certain window of parameters. The latter is presumably a more general feature due to the weakness of the binary-solute interaction. This scenario is in clear contrast with the effects generated by surfactants.

\begin{figure}[htbp]
\begin{center}
\centerline{\includegraphics[width = .46 \textwidth]{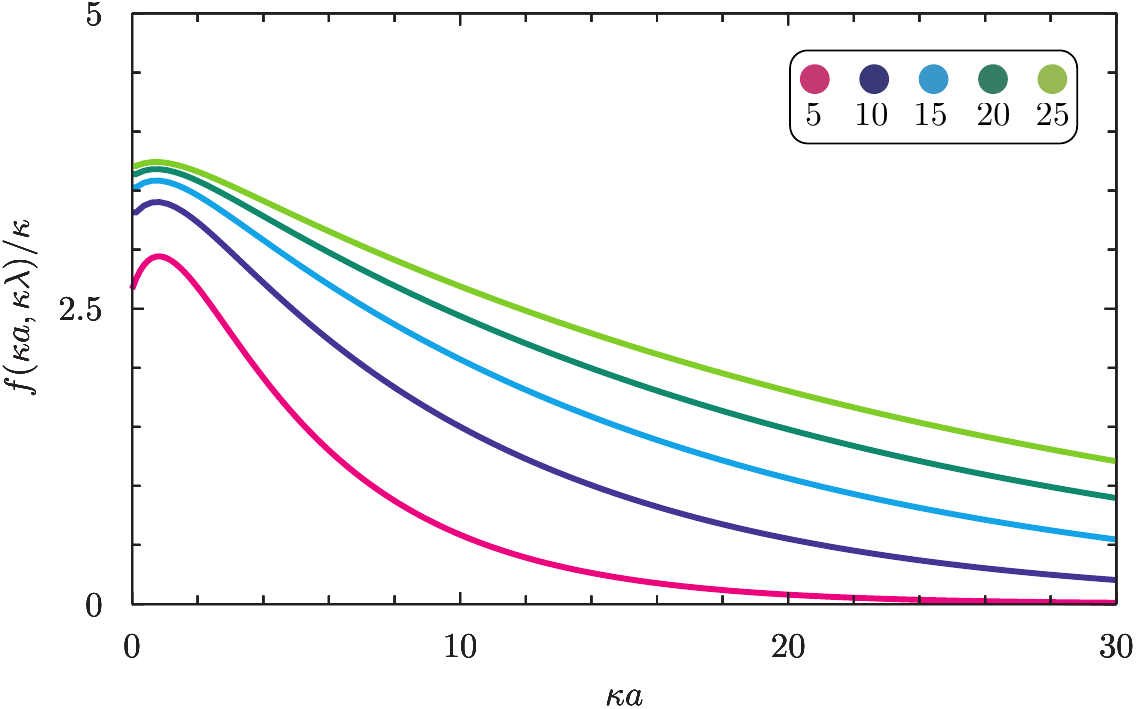}}
\caption{The rescaled force given by Eq.~(\ref{force}). The curves correspond to different values of the rescaled penetration depth $\kappa \lambda=5,10,15,20,25$ as indicated in the legend.}
\label{fig_force}
\end{center}
\end{figure}

We conclude this section by examining the force required to be exerted on the source-sink doublet in order to keep them in equilibrium. The free energy of the system is $\mathcal{F}=K\rho^{2} \mathcal{H}[\phi_{\Xi};\varrho]$. By using Eq. (\ref{A10}) and keeping terms up to $O(\bar{\omega})$, we have
\begin{equation}
\begin{aligned}
\label{newFE}
\mathcal{F} & = \gamma_{0} \mathcal{A}_{\pi} + \gamma_{1}[\varrho] \mathcal{A}_{\pi} + \gamma_{r}[\varrho] \mathcal{A}_{\pi} \, ,
\end{aligned}
\end{equation}
with $\gamma_{r}$ given by Eq. (\ref{A08}). It is convenient to isolate those terms which depend on the source, thus we write $\mathcal{F} = \gamma_{0} \mathcal{A}_{\pi} \Bigl[ 1 + gF \Bigr]$ with $F=(\gamma_{1}[\varrho]+\gamma_{r}[\varrho])/\gamma_{0}$, hence the interesting piece to analyze is $F$. We have the scaling form analogous to (\ref{scalingdouble})
\begin{equation}
\begin{aligned}
F(\kappa a,\kappa\lambda) & = \textrm{sign}(a) \biggl\{ -4 \kappa \lambda + \frac{ 4 \kappa \lambda }{ \kappa^{2} \lambda^{2}-1 } \biggl[ - 2 \textrm{e}^{-\kappa |a|} \\
& + (1+\kappa^{2}\lambda^{2}) \textrm{e}^{-|a|/\lambda} \biggr] \biggr\} \, .
\end{aligned}
\end{equation}
The function $F(\kappa a,\kappa\lambda)$ is odd with respect to $a$ and it is characterized by a sigmoidal profile which interpolates between $\pm 4\kappa\lambda$ for $a \rightarrow \mp \infty$. Since $F(\kappa a,\kappa\lambda)$ is a monotonous function, it reaches its minimum value for $\kappa a \rightarrow +\infty$. For $g>0$ the free energy is minimum when $a \rightarrow +\infty$, while for $g<0$ the free energy is minimum when $a \rightarrow -\infty$. The above considerations lead us to conclude that the source-sink doublet needs an external force $f$ in order to keep a constant position. The force turns out to be given by
\begin{equation}
\label{force}
f(\kappa a,\kappa\lambda) = - \partial_{a} F(\kappa a,\kappa\lambda) \, ,
\end{equation}
hence
\begin{equation}
\begin{aligned}
f(\kappa a,\kappa\lambda)/\kappa & = \frac{ 4 }{ \kappa^{2} \lambda^{2}-1 } \biggl[ - 2 \kappa \lambda \textrm{e}^{-\kappa |a|} \\
& + (1+\kappa^{2}\lambda^{2}) \textrm{e}^{-|a|/\lambda} \biggr] \, .
\end{aligned}
\end{equation}
Since $F(\kappa a,\kappa\lambda)$ decreases monotonically upon increasing $a$, the force $f(\kappa a,\kappa\lambda)$ is always positive; clearly it is an even function of $a$; see Fig. \ref{fig_force}. The force attains a certain nonzero value when $a=0$, it decreases for $\kappa |a| \gg1$ and reaches the maximum value for $\kappa a \approx 1$. Let us take $g>0$, meaning that the solute is produced in the less dense phase of the binary and it is destroyed in the denser phase of the binary. The force turns out to be positive, thus the separation between source and sink tends to increase.

\section{Local surface tension}\label{sec5}
So far we have considered an external field whose intensity $\varrho(z)$ does not change along the longitudinal directions parallel to the interface. Here we go one step beyond this hypothesis by considering an external field which may weakly change along the directions parallel to the interface, i.e., $x$ and $y$. In order to get analytical insight, we assume a slow variation of $\varrho$ along the parallel direction which occurs on a length scale $\lambda_{\|}$ much larger than the penetration length $\lambda$, hence $\lambda_{\|}\gg \lambda$.

Accordingly, the Laplacian operator in Eq.~(\ref{42}) can be decomposed as follows $\nabla^{2}=\nabla_{\|}^{2}+\partial_{z}^{2}$. The separation of scales we are assuming implies
\begin{equation}
| \nabla_{\|}^{2} \varrho | \approx \epsilon^{2} | \partial_{z}^{2} \varrho | \, .
\end{equation}
We seek for a solution of Eq.~(\ref{42}) in the form of an expansion in the small parameter $\epsilon \equiv \lambda/\lambda_{\|}$, hence we write
\begin{equation}
\varrho_{0}(\textbf{x}_{\|},z) = \sum_{k=0}^{\infty} \epsilon^{k} \varrho_{0}^{(k)}(\textbf{x}_{\|},z) \, .
\end{equation}
Due to the explicit dependence on the coordinates $\textbf{x}_{\|}$, we cannot any longer factor out the cross sectional area, therefore the free energy reads
\begin{equation}
\label{120}
\mathcal{F}[\varrho] = \iint_{\ell}\textrm{d}\textbf{x}_{\|} \, \gamma[\varrho^{(0)}(\textbf{x}_{\|},z)] \, ,
\end{equation}
where $\gamma[\varrho]$ is the free energy per unit area, as defined by Eq.~(\ref{newFE}), i.e., $\mathcal{F}/\mathcal{A}_{\pi}$ for the translationally invariant case. Accordingly, we have
\begin{equation}
\label{120-1}
\gamma[\varrho^{(0)}(\textbf{x}_{\|},z)] = \gamma_{0} + \widetilde{\gamma}_{1} [\varrho^{(0)}(\textbf{x}_{\|},z)] + O(\bar{\omega}^{2}) \, ,
\end{equation}
where $\widetilde{\gamma}_{1}[\varrho] = \gamma_{1}[\varrho] + \gamma_{r}[\varrho]$.

For a mildly undulated interface - thus in absence of overhangs - it is possible to use the Monge parametrization. The latter allows us to describe the interface in terms of the height function $z=\ell(\textbf{x}_{\|})$, thus the area $\mathcal{A}_{\ell}[\varrho]$ of the interface is given by
\begin{equation}
\mathcal{A}_{\ell}[\varrho] = \iint_{\ell}\textrm{d}\textbf{x}_{\|} \sqrt{ 1 + \left( \nabla\ell(\textbf{x}_{\|}) \right)^{2} } \, .
\end{equation}
For weak curvatures, the latter can be approximated by means of the square-gradient term
\begin{equation}
\label{121}
\mathcal{A}_{\ell}[\varrho] \approx \mathcal{A}_{\pi} + \frac{1}{2} \iint_{\ell}\textrm{d}\textbf{x}_{\|} \, \left( \nabla_{\|}\ell(\textbf{x}_{\|}) \right)^{2} \, ,
\end{equation}
where in the last line we exploit the smallness of the external field which allows to expand the square root in Eq.~(\ref{121}) and identify the first term as the area $\mathcal{A}_{\pi}$ for a flat interface. Therefore, up to quadratic corrections in the radii of curvature of the interface, the free energy per unit cross sectional area is now given by the ratio $\mathcal{F}/\mathcal{A}_{\pi}$. The above result may be interpreted as a \emph{local surface tension}; the latter can be identified with
\begin{equation}
\label{e87lst}
\gamma_{\textrm{loc}}(\textbf{x}_{\|}) = \gamma[\varrho^{(0)}(\textbf{x}_{\|},z)] \, ,
\end{equation}
where $\gamma[\varrho]$ is the functional given by Eq.~(\ref{120-1}). Keeping terms up to $O(\bar{\omega})$ in the interaction strength, Eq.~\eqref{e87lst} reads 
\begin{equation}
\gamma_{\textrm{loc}}(\textbf{x}_{\|}) = \gamma_{0} + \widetilde{\gamma}_{1}[\varrho^{(0)}(\textbf{x}_{\|},z)] + O(\bar{\omega}^{2})\, .
\end{equation}
The functional $\widetilde{\gamma}_{1}[\varrho]$ encodes the nontrivial spatial dependence of the local surface tension.

\subsection{At and away from equilibrium}
The notion of local surface tension provided by Eq.~\eqref{e87lst} needs to be taken carefully. In order to understand its implications, we will discuss the equilibrium and the non equilibrium settings.

\emph{Away from equilibrium}. 

Within the assumption of local equilibrium for the binary mixture, the binary liquid relaxes on time scales which are fast compared to the dynamics of the solute. The equality of generalized chemical potentials, i.e., $\Delta\mu_{\phi}(\textbf{r})=0$, is used in order to find the density of the binary liquid. In particular, if the solute density is invariant under translations along the interface ($\varrho=\varrho(z)$), then the chemical potentials of both components of the binary liquid are equal across the interface, thus $\mu_{a}(z)=\mu_{b}(z)$. On the other hand, if the solute density $\varrho(\bm{x}_{\|},z)$ breaks the invariance under translations along the interface, then the chemical potentials of the $a$ and $b$ components may acquire a dependence on the coordinates $\bm{x}_{\|}$, but still within the assumption of local equilibrium
\begin{equation}
\label{difference1}
\Delta\mu_{\phi}(\bm{x}_{\|},z) = \mu_{a}(\bm{x}_{\|},z)-\mu_{b}(\bm{x}_{\|},z)=0 \, .
\end{equation}
The latter means that for every liquid column with cross sectional area $\textrm{d}x\textrm{d}y$ centered in $z$, the system is at local phase coexistence.

\emph{At equilibrium}. Being at mechanical equilibrium, the sum of all stresses at the interface must vanish. De facto, Eq.~(\ref{e87lst}) supports the existence of non-vanishing local surface tension gradients at equilibrium with and external field. On such a basis one would conclude that the resulting inhomogeneous local surface tension will produce tangential stresses which will move the fluid, and eventually they lead to an incompatibility with the mechanical equilibrium. Although the local surface tension given by Eq.~(\ref{e87lst}) is inhomogeneous, the corresponding Marangoni stresses have to be inevitably compensated. As we are going to show, such a compensation requires a non-trivial condition for solute density.

At equilibrium we further require the chemical potentials $\mu_{a}$ and $\mu_{b}$ to be equal, but also their gradient along the interface has to be identical. Therefore Eq.~(\ref{difference1}) is supplemented with
\begin{equation}
\label{difference2}
\nabla_{\|} \Delta\mu_{\phi}(\bm{x}_{\|},z) = \textbf{0} \, .
\end{equation}
The solute density has to satisfy Eq.~(\ref{difference2}) in order to have absence of flows at equilibrium.

For the sake of completeness quote the expressions for the individual chemical potentials of both the binary liquid components
\begin{equation}
\begin{aligned}
\label{chpotentials}
\beta \mu_{a}(\textbf{r}) & = W(\phi) + \partial_{\phi}W(\phi) (\rho_{b}/\rho) - \frac{K}{2} \nabla^{2}\rho_{a} + \omega_{a}\varrho \, , \\
\beta \mu_{b}(\textbf{r}) & = W(\phi) - \partial_{\phi}W(\phi) (\rho_{a}/\rho) - \frac{K}{2} \nabla^{2}\rho_{b} + \omega_{b}\varrho \, .
\end{aligned}
\end{equation}
We conclude by recalling that for the $\phi^{4}$ model, the potential $W(\phi)$ is given by Eq.~(\ref{scaling}) with 
\begin{equation}
\label{phi4}
U(\phi) = \frac{\kappa^{2}}{8\phi_{0}^{2}} \left( \phi^{2}-\phi_{0}^{2} \right)^{2} \, ,
\end{equation}
$\phi_{0}$ given by Eq.~(\ref{bin007_a}) and $K\rho\kappa^{2} = 4(\chi-2)$. For the double-parabola model instead one uses Eq.~(\ref{potentialdp}) where $\kappa$ is the same as in Eq.~(\ref{phi4}).


\section{Conclusions}\label{sec:concl}
We have analyzed the dependence of the free energy and of the surface tension of a fluid interface on the density of a third suspended phase.
By means of a simplified model for the free energy of the unperturbed interface and via a perturbative approach we have derived a close formula for both the free energy and the surface tension. 

First, we have specialized to the case in which the density of the suspended phase is homogeneous along the interface. In this scenario our results show that, even at lowest order in the perturbative approach, the surface tension is sensitive to the presence of a third suspended phase. In particular, for the case-model that we have studied, our results predict a non-monotonous dependence of the surface tension on the distance of the source of the suspended phase. 

Second, when the density of the suspended phase is not homogeneous along the interface, our results predict a dependence of the surface tension on the local value of the density of the suspended phase. If the systems is at equilibrium, i.e. such an inhomogeneity is due to some external field, then the local force associated with the inhomogenous  surface tension will be balanced such that there will be no flow. In contrast, if the inhomogenous density of the suspended phase occurs in a non-equilibrium scenario, then the gradient of the surface tension will lead to the onset of Marangoni flows~\cite{Dominguez2016}.

Finally, our results can open the route to control diverse phenomena. For example, if catalytic colloids accumulate at one edge of the droplet, the imbalance in the density profile they generate will induce a net Marangoni flow that can induce net motion of the droplet~\cite{NAGAI2007,Chen2009,Frenkel2017}. Similarly, in the case in which the suspended phase attaches to the solid substrate it will induce an inhomogeneous solid-liquid surface tension that can pull droplets uphill~\cite{Chaudhury1539}. the stability of Pickering emulsions  upon tuning the release of an additional suspended phase.

\begin{acknowledgments}
AS and PM acknowledge A. Parry and J.-M. Romero Enrique for useful discussions.
\end{acknowledgments}


\appendix

\section{Free energy functionals}
\label{appendix}
Here we show how to derive Eq. (\ref{def203}) from Eq. (\ref{efe4prime}). Equation (\ref{efe4prime}) can be written as follows
\begin{equation}
\label{A01}
\delta^{\prime} \mathcal{F} = K\rho^{2} \mathcal{H}[\phi_{\Xi}(z; \varrho(z));\varrho(z)] - K\rho^{2} \mathcal{H}[\phi_{\rm sharp};\varrho(z)] \, .
\end{equation}
In order to compute the first addend in the right hand side of Eq. (\ref{A01}), we split the integral as follows $\int_{-\infty}^{+\infty}\textrm{d}z = \int_{-\infty}^{\ell}\textrm{d}z + \int_{\ell}^{+\infty}\textrm{d}z $, hence, by using the compact notation $\phi_{\Xi} \equiv \phi_{\Xi}(z;\varrho)$, we find
\begin{equation}
\begin{aligned}
\label{A02}
& \mathcal{H}[\phi_{\Xi}(z; \varrho(z));\varrho(z)] / \mathcal{A}_{\pi} = \\
& = \int_{-\infty}^{\ell} \biggl[ \frac{1}{2} \left( \partial_{z} \phi_{\Xi} \right)^{2} + \frac{ \kappa^{2} }{ 2 } (\phi_{\Xi}-\phi_{0})^{2} + \bar{\omega} \phi_{\Xi} \varrho\biggr] \\
& + \int_{\ell}^{+\infty} \biggl[ \frac{1}{2} \left( \partial_{z} \phi_{\Xi} \right)^{2} + \frac{ \kappa^{2} }{ 2 } (\phi_{\Xi}+\phi_{0})^{2} + \bar{\omega} \phi_{\Xi} \varrho\biggr] \, .
\end{aligned}
\end{equation}
In the first/second integral we can replace the square gradient with
\begin{equation}
\begin{aligned}
\label{A03}
& \left( \partial_{z} \phi_{\Xi} \right)^{2} = \partial_{z} \Bigl[ (\phi_{\Xi} \mp \phi_{0}) \partial_{z} \phi_{\Xi} \Bigr] - (\phi_{\Xi} \mp \phi_{0}) \partial_{z}^{2} \phi_{\Xi} \, ;
\end{aligned}
\end{equation}
being a total derivative, the first term contributes only through its evaluation at the boundaries, i.e., right at the interface, where $z \rightarrow \ell^{\pm}$, while at $z \rightarrow \pm \infty$ the contribution vanishes. Since we are evaluating the functional $\mathcal{H}[\phi_{\Xi};\varrho]$ for the profile $\phi_{\Xi}$ which satisfies the associated mean-field (Euler-Lagrange) equations (see Eq. (\ref{C_03})), we can use Eq. (\ref{C_03}) in order to express $\partial_{z}^{2} \phi$ in Eq. (\ref{A03}); a simple calculation yields
\begin{equation}
\begin{aligned}
\label{A04}
& \mathcal{H}[\phi_{\Xi}(z; \varrho(z));\varrho(z)] / \mathcal{A}_{\pi} = \\
& = - \frac{ \phi_{0}}{2} \biggl[ \partial_{z}\phi_{\Xi}(z;\varrho)\big\vert_{z=\ell^{-}} + \partial_{z}\phi_{\Xi}(z;\varrho) \big\vert_{z=\ell^{+}} \biggr] \\
& + \frac{ \bar{\omega} }{2} \int_{-\infty}^{\ell}\textrm{d}z \, \varrho(z) [ \phi_{\Xi}(z;\varrho) + \phi_{0} ] \\
& + \frac{ \bar{\omega} }{2} \int_{\ell}^{\infty}\textrm{d}z \, \varrho(z) [ \phi_{\Xi}(z;\varrho) - \phi_{0} ] \, .
\end{aligned}
\end{equation}
In the above we have also used the crossing criterion, i.e., $\phi_{\Xi}(\ell)=0$. Now we use Eq. (\ref{C_10}), recall that $\Delta\phi_{\Xi}(z;\varrho_{0})=O(\bar{\omega})$, and collect terms up to $O(\bar{\omega})$; we have
\begin{equation}
\begin{aligned}
\label{A05}
& \mathcal{H}[\phi_{\Xi}(z; \varrho(z));\varrho(z)] / \mathcal{A}_{\pi} = \\
& = - \frac{ \phi_{0}}{2} \biggl[ \partial_{z}\phi_{\Xi}(z;0)\big\vert_{z=\ell^{-}} + \partial_{z}\phi_{\Xi}(z;0) \big\vert_{z=\ell^{+}} \biggr] \\
& - \frac{ \phi_{0}}{2} \biggl[ \partial_{z}\Delta\phi_{\Xi}(z;\varrho_{0})\big\vert_{z=\ell^{-}} + \partial_{z}\Delta\phi_{\Xi}(z;\varrho_{0}) \big\vert_{z=\ell^{+}} \biggr] \\
& + \frac{ \bar{\omega} }{2} \int_{-\infty}^{\ell}\textrm{d}z \, \varrho_{0}(z) [ \phi_{\Xi}(z;0) + \phi_{0} ] \\
& + \frac{ \bar{\omega} }{2} \int_{\ell}^{\infty}\textrm{d}z \, \varrho_{0}(z) [ \phi_{\Xi}(z;0) - \phi_{0} ] \\
& + O(\bar{\omega}^{2}) \, .
\end{aligned}
\end{equation}
Notice that in the last two terms we have used the unperturbed density profile $\phi_{\Xi}(z;0)$ and considered the zeroth-order solution $\varrho_{0}(z)$ because we are interested to work out terms up to order $O(\bar{\omega})$.

We can easily see that for $\varrho=0$ the only term which survives in the right hand of Eq. (\ref{A05}) is the first one, the latter is actually the surface tension $\gamma_{0}$ of an interface in the absence of the external field, i.e.,
\begin{equation}
\label{A06}
\frac{ \gamma_{0} }{ K\rho^{2} } = - \frac{ \phi_{0}}{2} \biggl[ \partial_{z}\phi_{\Xi}(z;0)\big\vert_{z=\ell^{-}} + \partial_{z}\phi_{\Xi}(z;0) \big\vert_{z=\ell^{+}} \biggr] \, ,
\end{equation}
a result which follow straightforwardly from the interface profile $\phi_{\Xi}(z;0)$ given by Eq. (\ref{profile_DP}). Coming back to (\ref{A05}), inside the third integral we can use $\phi_{\Xi}(z;0) = \phi_{0} - \phi_{0}\mathcal{K}(z;\ell)$ while in the fourth one we can use $\phi_{\Xi}(z;0) = - \phi_{0} + \phi_{0}\mathcal{K}(z;\ell)$, thus
\begin{equation}
\begin{aligned}
\label{A07}
& \mathcal{H}[\phi_{\Xi}(z; \varrho(z));\varrho(z)] / \mathcal{A}_{\pi} = \\
& = \frac{ \gamma_{0} }{ K\rho^{2} }   - \frac{ \phi_{0}}{2} \biggl[ \partial_{z}\Delta\phi_{\Xi}(z;\varrho_{0})\big\vert_{z=\ell^{-}} + \partial_{z}\Delta\phi_{\Xi}(z;\varrho_{0}) \big\vert_{z=\ell^{+}} \biggr] \\
& - \frac{ \bar{\omega} \phi_{0} }{ 2 } \int_{-\infty}^{\ell}\textrm{d}z \, \varrho_{0}(z) \mathcal{K}(z,\ell) + \frac{ \bar{\omega} \phi_{0} }{ 2 } \int_{\ell}^{\infty}\textrm{d}z \, \varrho_{0}(z) \mathcal{K}(z,\ell) \\
& + \bar{\omega} \int_{-\infty}^{\infty}\textrm{d}z \, \varrho_{0}(z) \phi_{\Xi}(z;0)
+ O(\bar{\omega}^{2}) \, .
\end{aligned}
\end{equation}
It is then a simple matter to show that the second term in the right hand side of Eq. (\ref{A07}) coincides with the sum of the two integrals in the second line of the right hand side of of Eq. (\ref{A07}). Moreover, it is immediate to show that the last integral in Eq. (\ref{A07}) coincides with the reference free energy per unit area
\begin{equation}
\begin{aligned}
\label{A08}
\gamma_{r}[\varrho] & = K\rho^{2} \mathcal{H}[\phi_{\rm sharp};\varrho(z)]/\mathcal{A}_{\pi} \\
& = K\rho^{2} \bar{\omega} \int_{-\infty}^{\infty}\textrm{d}z \, \varrho_{0}(z) \phi_{\Xi}(z;0) \, .
\end{aligned}
\end{equation}
Now we introduce
\begin{equation}
\begin{aligned}
\label{A09}
\gamma_{1}[\varrho] & = - K\rho^{2} \bar{\omega} \phi_{0} \int_{-\infty}^{\ell}\textrm{d}z \, \varrho_{0}(z) \mathcal{K}(z,\ell) \\
& + K\rho^{2} \bar{\omega} \phi_{0} \int_{\ell}^{\infty}\textrm{d}z \, \varrho_{0}(z) \mathcal{K}(z,\ell) \, ,
\end{aligned}
\end{equation}
and thus Eq. (\ref{A07}) is written in the compact form
\begin{equation}
\begin{aligned}
\label{A10}
K\rho^{2} \mathcal{H}[\phi_{\Xi};\varrho] / \mathcal{A}_{\pi} & = \gamma_{0} + \gamma_{1}[\varrho] + \gamma_{r}[\varrho] + O(\bar{\omega}^{2}) \, ,
\end{aligned}
\end{equation}
which implies for Eq. (\ref{A01}) 
\begin{equation}
\label{A11}
\delta^{\prime}\mathcal{F} = \gamma_{0} \mathcal{A}_{\pi} + \gamma_{1}[\varrho] \mathcal{A}_{\pi} + O(\bar{\omega}^{2}) \, .
\end{equation}

\bibliographystyle{unsrt}
\bibliographystyle{apsrev4-1}
\bibliography{bibliography}

\end{document}